\documentclass[runningheads]{svmult}

\usepackage{makeidx}   % allows index generation
\usepackage{graphicx}  % standard LaTeX graphics tool
\usepackage{subeqnar}  % subnumbers individual equations
\usepackage{multicol}  % used for the two-column index
\usepackage{physprbb}  % modified textarea for proceedings,
\usepackage{float}
\makeindex             % used for the subject index
\begin{document}
\title*{GRB~000301C  : a   possible  short/intermediate duration burst
connected to a DLA system}
\toctitle{The optical afterglow of GRB~000301C}
% allows explicit linebreak for the table of content
%
%
\titlerunning{The optical afterglow of GRB~000301C}
% allows abbreviation of title, if the full title is too long
% to fit in the running head
%
\author{J. Gorosabel\inst{1,2}
\and J.U. Fynbo\inst{2}
\and B.L. Jensen\inst{3}
\and P. M\o ller\inst{2}
\and H. Pedersen\inst{3}
\and J. Hjorth\inst{3}
\and M. I. Andersen \inst{4}
\and K. Hurley \inst{5}}
\authorrunning{J. Gorosabel et al.}

\institute{Danish Space Research Institute, Juliane Maries Vej 30,
DK--2100 Copenhagen \O, Denmark\\
\and European Southern Observatory, Karl--Schwarzschild--Stra\ss e 2,
D--85748 Garching, Germany\\
\and Astronomical Observatory, University of Copenhagen,
Juliane Maries Vej 30, DK--2100 Copenhagen \O, Denmark
\and Division of Astronomy, P.O. Box 3000, FIN-90014 University of Oulu,
Finland.
\and Space Science Laboratory, University of California at Berkerley, USA.
}

\maketitle              % typesets the title of the contribution

\begin{abstract} 

We    discuss    two  main  aspects   of    the  GRB~000301C afterglow
\cite{Fynb00,Jens01}; its short  duration and  its possible connection
with  a  Damped Ly$\alpha$  Absorber (DLA).   GRB~000301C falls in the
short class of bursts,  though it is consistent  with belonging to the
proposed   intermediate    class or  the extreme   short    end of the
distribution of long-duration  GRBs.  Based  on   two VLT spectra   we
estimate  the   \rm{H   I} column   density    to   be  $\log$(N(\rm{H
I}))$=21.2\pm0.5$.    This is  the first    direct   indication of   a
connection between GRB host galaxies and Damped Ly$\alpha$ Absorbers.
\end{abstract}

\section{Introduction}

GRB~000301C was  localised by  the  Inter Planetary Network (IPN)  and
RXTE to an area of $\sim$50 arcmin$^2$.   A fading optical counterpart
was subsequently  discovered  with the Nordic Optical  Telescope (NOT)
about 42~h after  the burst.  The GRB was  recorded by the Ulysses GRB
experiment and by the  NEAR $X$-Ray/Gamma-Ray Spectrometer.   From the
NEAR data we estimate the 150--1000~keV fluence to be approximately $2
\times 10^{-6}\, \mathrm{erg\, cm}^{-2}$.   The IPN/RXTE  error-box of
GRB~000301C \cite{SHC2000} was observed with the 2.56-m Nordic Optical
Telescope (NOT) on 2000 March 3.14--3.28  UT ($\sim$1.8 days after the
burst) using ALFOSC.  Comparing with  red and blue Palomar Optical Sky
Survey~II exposures,  a candidate Optical Transient  (OT) was found at
the  position $(\alpha,\delta)_{2000}$=($16^h$     $20^m$   $18.56^s$,
$+29^{\circ}$ $26^{\prime}$ $36.1^{\prime \prime})$.

\section{The first short GRB optical counterpart detection?}

\begin{figure}[H]
\begin{center}
\resizebox{11.15cm}{!}{\includegraphics{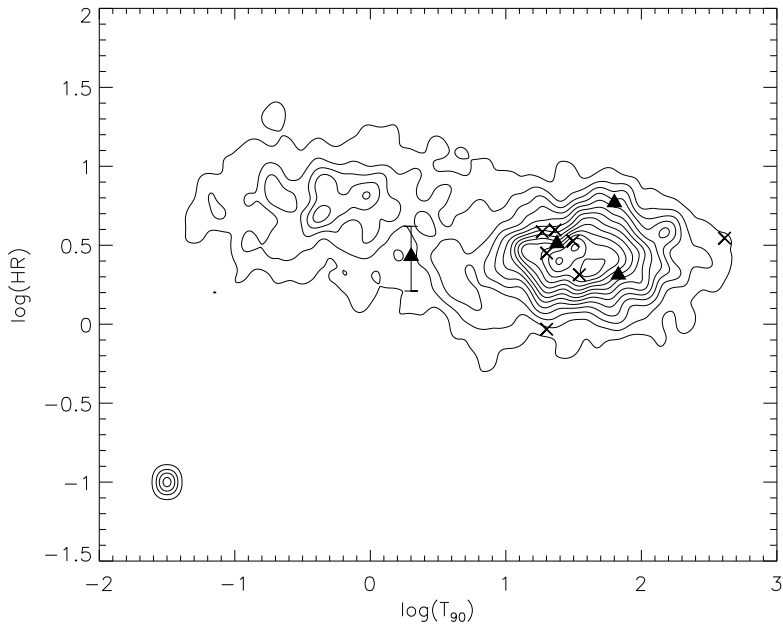}}
\end{center}
\caption[]{A contour plot showing the duration--hardness
($\log$(T$_{90}$)--$\log$(HR$_{32}$)) distribution of 1959 BATSE 
bursts.  The triangle with an error-bar near the center of the plot 
represents GRB~000301C. As seen, the burst was located in the 
short/intermediate duration part of the distribution. Other symbols 
represent 10 other BATSE bursts with identified optical counterparts 
for which data on fluence and duration
are available. Triangles are bursts which have a break in their
optical light curves. Errors in the BATSE data are smaller than the
symbol size. Contour levels scale linearly. The point in the lower
left corner illustrates the resolution of the contours.}
\label{eps1}
\end{figure}

  As measured by both Ulysses and NEAR, in the $>$25~keV energy range,
the duration of this  burst was approximately  2 s.  GRB~000301C falls
in the short class of  bursts, though it  is consistent with belonging
to the proposed intermediate class  or the  extreme  short end of  the
distribution of  long-duration  GRBs.  We  obtain a hardness  ratio of
2.7$\pm$0.6(cutoff)$\pm$30\%(statistical  error, see \cite{Jens01} for
details on the  calculation  of the hardness  ratio).  Fig.~\ref{eps1}
shows the location of GRB~000301C in a hardness vs. duration plot.

\section{The first GRB-DLA connection?}

\begin{figure}[H]
\resizebox{9.5cm}{!}{\includegraphics[angle=270]{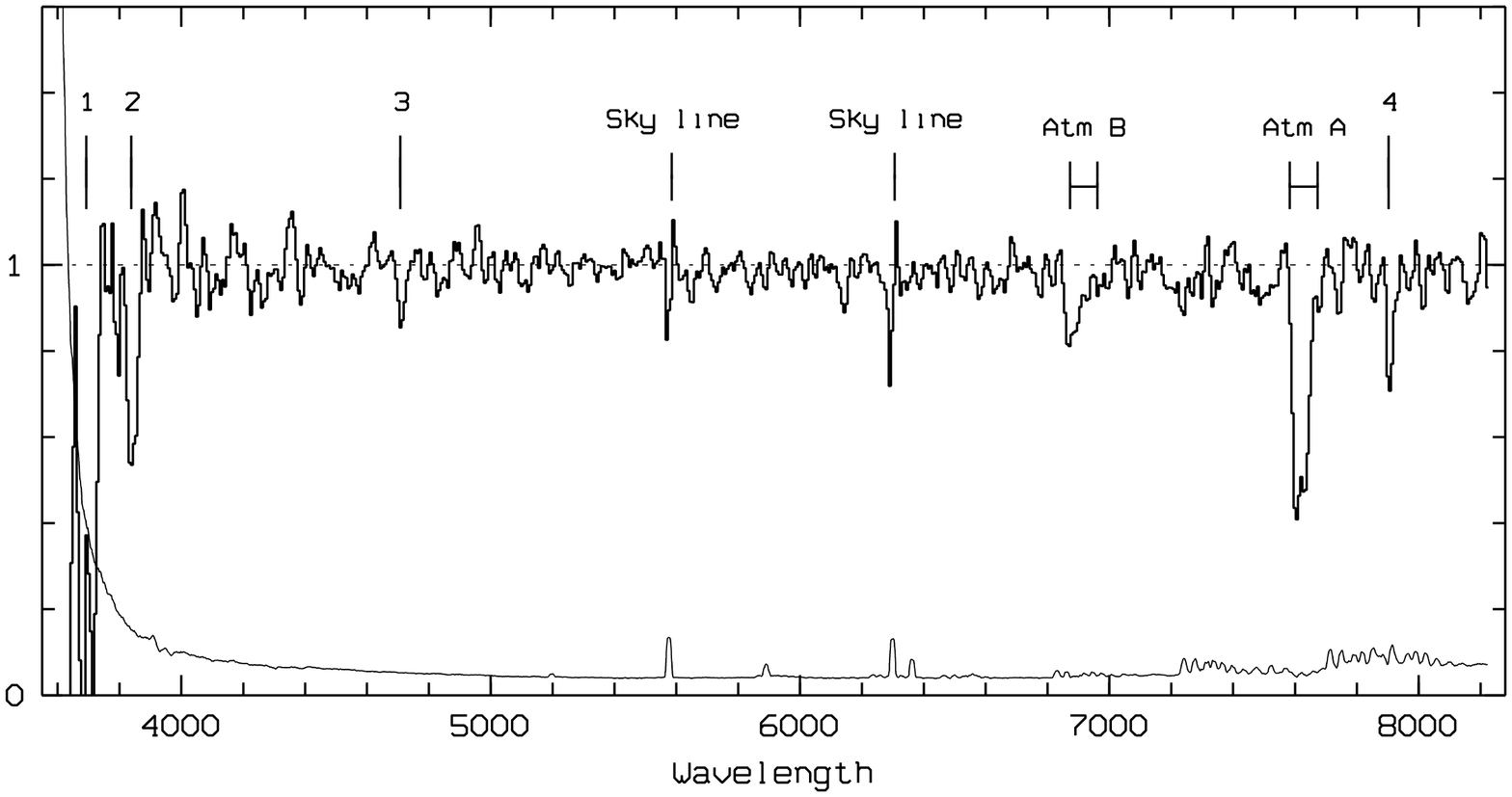}}
\resizebox{2.2cm}{!}{\includegraphics[angle=270]{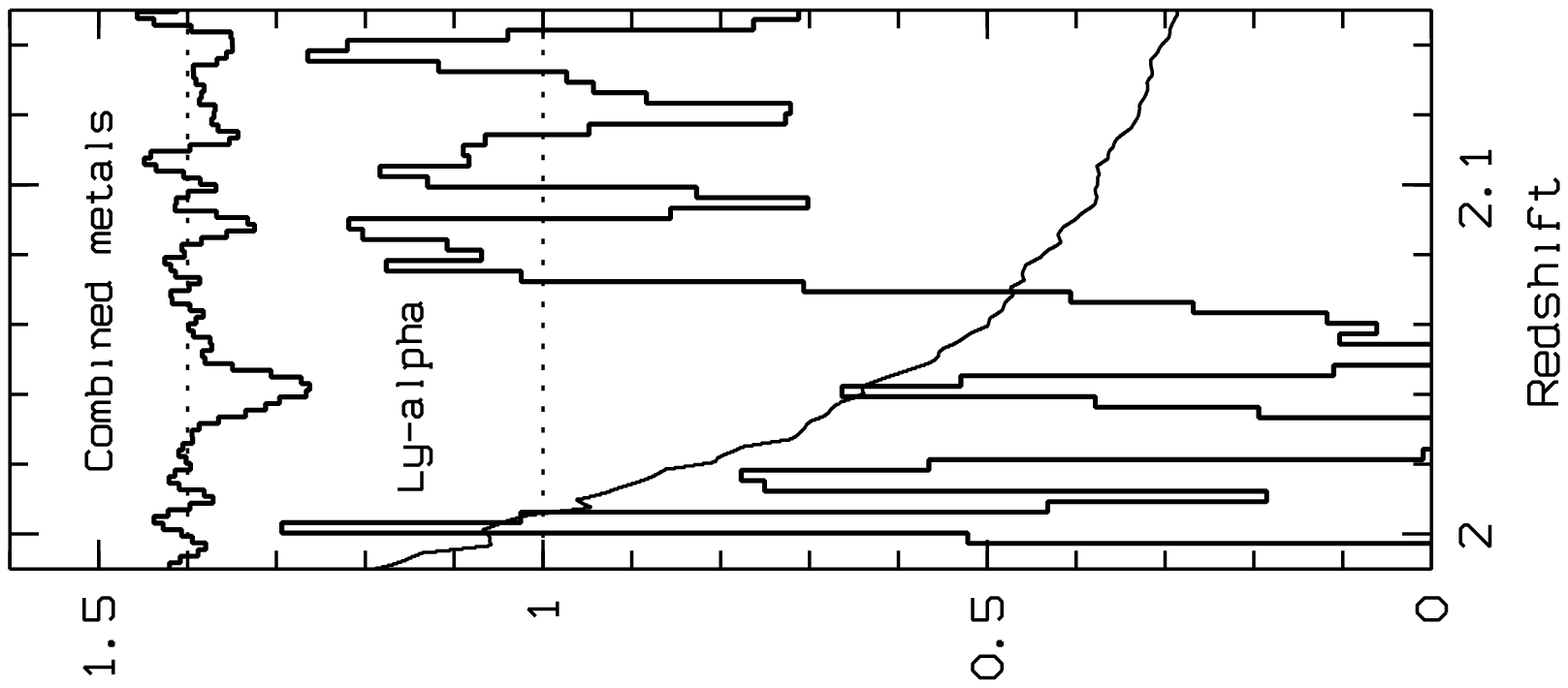}}
\caption{Combined VLT+FORS1 spectrum of   GRB~000301C from 2000  March
5+6  UT\@.  The spectrum is normalized  to  1 in  the  continuum.  The
atmospheric absorption bands and residuals  from strong sky--lines are
marked, as well as the 4 absorption lines. The spectrum is binned to 7
{\AA} pixels, and the  lower curve shows  the noise (per pixel).   The
detected lines are: Ly$\alpha$ (1), SI  \rm{II} (2), C \rm{IV} (3) and
Fe   \rm{II} (4).  On the  right  side we  can see  a   blow up of the
Ly$\alpha$ region  in the  redshift  space.  We  have overplotted  the
oscillator  strength weighted mean  of metal  lines lines (\rm{Fe II},
\rm{Si II} and \rm{C  II}). The lower  curve shows the noise per  bin.
Note the   very sharp  onset of absorption   well above  the  expected
redshift.  This is consistent  with a very broad Ly$\alpha$ absorption
line as detailed in the text.  }
\label{FIGURE:Spectrum}
\end{figure}

Spectroscopic observations were carried out  on 2000 March  5 and 6 UT
with  VLT-Antu equipped  with FORS1.  Fig.~\ref{FIGURE:Spectrum} shows
the normalized spectrum of  the OT.  Following the procedure explained
in  \cite{Jens01} we obtained a redshift  of $z_{\rm abs} = 2.0404 \pm
0.0008$.  The oscillator strength   weighted mean observed  equivalent
width of the  Fe \rm{II} lines  is 2.56~{\AA}, which is strong  enough
that by comparison to known quasar absorbers one  would expect this to
likely have a column density of neutral Hydrogen in excess of $2\times
10^{20}   \mathrm{cm}^{-2}$.    Such absorbers  are  known   as Damped
Ly$\alpha$ Absorbers (DLAs),  and hold a  special interest because  of
the    large  amounts  of   cold  gas  locked  up    in  those objects
\cite{SLI1997}. On  the other  hand  we found that the  spectrum drops
steeply before the expected  central position of the Ly$\alpha$  line,
and well before the S/N drops below detection  (see right side plot of
Fig.~\ref{FIGURE:Spectrum}).   One  likely explanation for  this is the
pre\-sence of  a very broad Ly$\alpha$   absorption line.  To quantify
this  we have  modelled several  Ly$\alpha$  absorption lines,  all at
redshift 2.0404. The formal $\chi^2$ minimum is  found at N(H \rm{I})$
= 1.5\times 10^{21}  \mathrm{cm}^{-2}$ ($\chi^2$ per  DOF = 0.86), but
any value within a factor 3 of this is acceptable.

%=====================Begin Figure Spectrum===========================
%=====================End Figure Spectrum=============================

%Due to the poor resolution, only very strong absorption lines can be
%detected individually. In Table~\ref{TABLE:abs_lines} we list the only
%four absorption features which were detected at a S/N in excess of
%4.5. 

%%=====================Begin Table abs_lines==================================
%\begin{table}[t]
%\begin{center}
%\caption{Absorption features detected at S/N$>$4.5  in the spectrum of
%GRB~000301C on   2000 March  5+6  UT\@. W$_{\rm   obs}$  indicates the
%observer frame equivalent width of the features.}
%\label{TABLE:abs_lines}
%\begin{tabular}{lccrr}
%\hline
%Line & Identification  & $\lambda_{\rm vac}$ & W$_{\rm obs}$ & $\sigma$(W)\\
%                       &  ({\AA})            &  ({\AA})  &   &            \\
%\hline
%1 & Ly$\alpha$      & 3693.56 & 67.07 & 12.09 \\
%2 & SI \rm{II}~1260 & 3843.09 & 18.99 & 2.03  \\
%3 & C  \rm{IV}~1549 & 4712.23 &  3.76 & 0.82  \\
%4 & FE \rm{II}~2600 & 7909.03 &  6.29 & 1.21  \\
%\hline
%\end{tabular}
%\end{center}
%\end{table}
%%=====================End Table Redshift==================================
\section{Conclusions}
GRB~000301C is  so far the   GRB of shortest   duration, for which  a
counterpart   has been detected.   The   high-energy properties of the
burst are  consistent with membership  of the  short-duration class of
GRBs, though GRB~000301C  could  belong to the   proposed intermediate
class of  GRBs  or the   extreme  short end   of  the distribution  of
long-duration GRBs.   We argue that there  may be a connection between
the  host galaxy of  GRB~000301C and DLAs.

%INDEX%%%%%%%%%%%%%%%%%%%%%%%%%%%%%%%%%%%%%%%%%%%%%%%%%%%%%%%%%%%%%%%
% Please check with the editor of your book whether he plans to
% include a "mutual" subject index - if so, please code your entries
% in the standard syntax. For your own purposes you may print your
% "personal" index by using the following commands:
%
%\clearpage
%\addcontentsline{toc}{section}{Index}
%\flushbottom
%\printindex
%%%%%%%%%%%%%%%%%%%%%%%%%%%%%%%%%%%%%%%%%%%%%%%%%%%%%%%%%%%%%%%%%%%%%

\end{document}